\documentclass[12pt]{article}
\newcommand{\dpf}{\displaystyle\frac}
\title{{\small \rightline{\vbox{\hbox{\rightline{McGill/97-28}}
\hbox{\rightline{SUNY-NTG-97-49}}}}}
\ \\ Off-shell effects in dilepton production from hot interacting
mesons}

\author{Song Gao \thanks{Current address: Institut f\"ur Theoretische Physik 
der J. W. Goethe Universit\"at, Robert-Mayer Str. 10, Postfach 11 19 32,
D-60054 Frankfurt a.M., Germany.}
 \vspace*{0.1in} \\
{\it\small Physics Department, McGill University}\\ \vspace*{0.2in}
{\it\small 3600 University St., Montr\'eal, QC, H3A 2T8, 
Canada}\\  \vspace*{0.1in}    
Charles Gale\thanks{Permanent address:  Physics Department, 
McGill University, 3600 University St., Montr\'eal, QC, H3A 2T8, Canada.}
 \vspace*{0.1in} \\
{\it\small Department of Physics and Astronomy, 
SUNY at Stony Brook}\\ \vspace*{0.2in}
{\it\small Stony Brook, New York, 11794-3800}\\ \vspace*{0.1in}}

\date{ }
\usepackage[dvips]{graphicx}
\begin{document}
\maketitle
\begin{abstract}
    The production of dielectrons in reactions involving 
$a_{1}$ mesons  and pions is studied. We compare results obtained with 
different phenomenological Lagrangians that have been used 
in connection with 
hadronic matter and finite nuclei. We insist on the necessity for those
interactions to satisfy known empirical properties of the strong
interaction. Large off-shell effects in dielectron production 
are found and some consequences 
for the interpretation of heavy ion data are outlined. We also 
compare with results
obtained using experimentally-extracted spectral functions. 
\end{abstract}

%\pacs{PACS : 25.75.-q, 13.75.Lb, 12.38Mh}

%\vskip2pc]

\section{Introduction }

The field of heavy ion collisions is a very active one, straddling high
energy and nuclear physics. At the upper energy limit of this
flourishing area of research,  the goal is to eventually produce and
study a new
state of matter in the laboratory: the quark-gluon plasma (QGP). That 
strongly
interacting matter in conditions of extreme energy densities undergoes a phase
transition is in fact  a prediction of QCD \cite{qcd}. 
To confirm whether the phase transition indeed occurs in relativistic
heavy ion collisions and that the QGP is
formed, one needs a clear signal as a signature for the QGP.
Many approaches have been suggested to elucidate the existence of this
elusive state of matter, but unfortunately no single measurement can yet be
singled-out as a ``smoking gun'' candidate. Instead, it appears that  
many complementary
experimental data will require simultaneous analysis \cite{harris96}. 
One class of observables that appears especially attractive is that of
electromagnetic signals. This owes to the fact that such probes
essentially suffer little or no final state interactions and thus
constitute reliable carriers able to report on the local conditions at
their emission site.  Indeed, the calculated emission rates for 
photons and lepton
pairs have been shown to strongly depend on the local density and
temperature. Those facts were established some time ago and several
experiments specializing in the measurement of electromagnetic radiation
are either running right now or being planned. 

In this paper, we shall be concerned about lepton pairs and 
calculations of their production. With respect to photon measurements, 
they constitute an important related observable. The extra degree of
freedom associated with the invariant mass makes them valuable tools
in the study of annihilation reactions, for example. The emission of
lepton pairs from the QGP and also from hot hadronic has been studied by
many authors \cite{qm96}. We will not discuss here the lepton pairs with
high invariant masses ($M \geq  m_{J/\psi}$), they constitute a
fascinating story of their own and the literature on the subject is
considerable. Rather, we shall address the issue of softer pairs and
that of the reliability of calculations of their emission.  Recently, a
lot of attention has focused on measurements from the CERES/NA45
collaboration \cite{axel96}. Several theoretical calculations have
attempted to reproduce this experimental data \cite{axel96}. 
Because of the exciting potential of such
measurements to reveal new physics, extra care has to be taken in their 
theoretical modeling. 
In the very soft sector, relevant for the CERES
experiment, different theoretical calculations for the fundamental
processes have in fact been confronted wih each other with a striking
degree of agreement \cite{jaipur}. It has also been shown that this data
supports the fact that the very soft sector of the lepton pair spectrum is
relatively independent of the dynamical approach used to model the
evolution of the four-volume of the colliding system. However, the
disagreement between models and data has led to different
interpretations \cite{cerestheo}.

Another interesting aspect of the CERES analysis is that the region of
invariant mass  $m_\phi < M < m_{J/\psi}$ also shows an ``anomalous'' 
excess of
lepton pairs, when compared with estimates based on hadronic 
electromagnetic 
decays. Note that this excess is seen only in nucleus-nucleus events,
the proton-nucleus results are well reproduced assuming hadronic
sources \cite{axel96}. Adopting the same line of reasoning as 
that used in most 
of the calculations done to
interpret the low mass sector of the CERES experiment, it is natural to
ask which hadronic reactions could contribute significantly to the mass
region in question. Some work on this aspect was performed in the past 
where it was found that in a hot environment of mesons, the reaction $\pi
+ a_1 \rightarrow e^+ e^-$ should be an important source of dileptons
\cite{skg,kimko,hg} 
(from here on we shall discuss only dielectrons).
Such studies concerned with the contribution of the $a_1$ to 
the lepton pair spectrum
are based on
effective chiral Lagrangians containing $\rho$, $\pi$ and $a_1$. Our purpose
in this paper is the following: we review several different effective 
Lagrangians for the $\pi \rho a_1$ vertex, and we calculate lepton pair
production from the $\pi
+ a_1 \rightarrow e^+ e^-$ reaction, using the Vector Meson Dominance 
model (VMD).
We shall follow a philosophy similar to that in Ref. \cite{kimko}, but
we go further by including a survey of other known and used 
interactions. We shall 
also stress the importance of conforming to empirical 
measurements, and we highlight the off-shell properties. 
Our work is organized as follows: after introducing the different models
and mentioning their origins, 
we compare with each other the lepton results obtained and comment on the
restrictions imposed on the effective interactions by the 
hadronic phenomenology. We also will show results from an 
experimentally-extracted  spectral
function analysis of lepton production processes. We then conclude. 

\section{Effective Lagrangians and phenomenology of the $a_1$ meson}

In our survey of the $\pi \rho a_1$ interaction, we in turn will 
consider the following Lagrangians:
\begin{enumerate}
\item An effective chiral Lagrangian 
based on the $SU(2)_L \times SU(2)_R $
linear $\sigma$ model \cite{ko}.
\item A  $U(2)_L \times U(2)_R $ chiral model for pseudoscalar,
vector, and axial-vector mesons \cite{li}.
\item An effective chiral Lagrangian where the vector mesons are
introduced as massive Yang-Mills fields of the chiral symmetry 
\cite{gom}.
\item An effective Lagrangian previously used 
in connection with photon emission rates \cite{xiong}.
\item An effective Lagrangian used 
to address the issue of form factors in the Bonn potential
\cite{jhs}.
\end{enumerate}

As one can realize, many interactions claim to address the $a_1 \pi \rho$ 
vertex. This variety owes to the fact that there is currently no 
unique way of 
implementing the chiral symmetry in an approach uniting pseudoscalars, 
vectors and pseudovectors. Therefore, many versions of a given coupling can 
be obtained, with symmetry requirements sometimes yielding to 
specific practical concerns. Note that all 
of the above Lagrangians (except one) have been used in the literature to 
calculate the lepton-producing reaction we are considering here. In such 
investigations, a popular practice consists of adjusting any free 
parameter in the above
interactions to reproduce the experimental decay width  
$\Gamma_{a_1 \rightarrow \pi \rho}$. In cases where more
than one free parameter is available, one can also fit 
$\Gamma_{a_1 \rightarrow \pi \gamma}$, when electromagnetic phenomena 
are of some
relevance. 
However, another well-defined empirical quantity exists, 
regarding the decay $a_1
\rightarrow \pi \rho$. The ratio of $D$-wave to $S$-wave (eigenstates of
the relative orbital angular momentum in the exit channel) amplitude has been 
measured and is known to
be $D/S~=~-0.09~\pm~0.03$~\cite{pdg96}. Theoretically, the calculation of
this ratio is accomplished by expanding the decay amplitude in the helicity
basis, using the standard helicity representation for the polarization
vectors. On the other hand, one can expand the amplitude in spherical
harmonics. Working in the rest frame of the decaying particle, in this  case
the $a_1$, one can then match the two  representations in terms of their
helicity content. It is probably fair to say that this aspect of hadronic
phenomenology has not received the attention it should have in the heavy
ion community by 
users of effective hadronic Lagrangians.

We will now proceed as advertised earlier and compare in turn the
interactions described above in terms of their contribution to the 
process $\pi + a_1 \to e^+ e^-$. However, before we turn to this
specific application, it is instructive to recall why the process in
question is not determined uniquely by choosing an interaction and fitting
on-shell properties.  
For our
illustrative purpose, consider the process $a_1 \to \rho \pi$.  The most
general vertex for this strong decay can be written as
\begin{eqnarray}
\Gamma^{\mu \nu} \ = \ i g_1 \,g^{\mu \nu} + i g_2\, q^\mu k^\nu +
i g_3\, k^\mu q^\nu +
i g_4\, q^\mu q^\nu + i g_5\, k^\mu k^\nu\ ,
\label{vertex}
\end{eqnarray}
where $g_i (k^2, q^2)$ is a form factor, with
$k^\mu$ and $q^\nu$ being the the $a_1$ and $\rho$
four-momenta. For on-shell $a_1$ decay, three of the above terms are
identically zero, owing to the transversality condition 
$\epsilon (a_1) \cdot k = \epsilon (\rho) \cdot q = 0$.  The on-shell 
form factors are usually chosen
such that (some of) the experimental constraints discussed above are satisfied.
However in the reaction under scrutiny here, $\pi + a_1 \to \rho \to 
e^+ e^-$, the
$\rho$ meson is not on its mass shell and the extrapolations of form
factors to this region are not unique. To evaluate the importance of such
effects is the purpose of this work. Note however that off shell, for our 
reaction, there are still only two terms that are relevant
in Eq. (\ref{vertex}). This is because the lepton electromagnetic current
is conserved. 
More specifically, the other
term of Eq.~(\ref{vertex}) that could  in principle contribute to 
our reaction is 
$i g_4 q^\mu q^\nu$. However, the corresponding 
matrix element for electron-positron pair production, once the spin 
sums are carried out,  will look like (the lepton masses have been set to
zero, for simplicity) 
\begin{eqnarray}
|\overline{\cal M} |^2\ \propto\ q^\mu q^\nu {\rm Tr} \left( \gamma_\mu 
\not\! q_1 \gamma_\nu \not\!q_2\right)\ ,
\label{prob}
\end{eqnarray}
where $q_1^\mu (q_2^\mu)$ is the four-momentum of the electron (positron).
Note that now $q^\mu\ = \ q_1^\mu + q_2^\mu$. 
The above expression, Eq.~(\ref{prob}), is zero. 

\subsection{A Chiral $SU(2)_L \times SU(2)_R $ effective Lagrangian}

	The effective Lagrangian for the $a_1 \pi \rho$ vertex constructed by
Ko and Rudaz \cite{ko} is based on the $SU(2)_L \times SU(2)_R $
linear $\sigma$ model, with the $\rho$ vector meson and the $a_1$ axial-vector
meson included as
phenomenological gauge fields associated with a local  $SU(2)_L
\times SU(2)_R $ chiral symmetry. This local chiral symmetry is then broken
to global chiral symmetry. 

     The $a_1 \rho \pi$ vertex is generated by the following interaction
Lagrangian\cite{kimko,ko},
\begin{eqnarray}
{\cal L}_{a_{1} \rho \pi} & = & {g^{2} f_{\pi} \over Z_{\pi}}~\left[ 2 c
\vec{\pi} \cdot ( \vec{\rho}_{\mu} \times \vec{a}^{\mu} ) +
{1 \over 2 m_{a_1}^2}~ \vec{\pi} \cdot ( \vec{\rho}_{\mu\nu}
\times \vec{a}^{\mu\nu} )      \right.
\nonumber    \\
& & +  \left. \left( { 1\over m_{a_1}^2} - {\kappa_{6} Z_{\pi}
\over m_{\rho}^2} \right) \partial_{\mu} \vec{\pi} \cdot (
\vec{\rho}^{\mu \nu} \times \vec{a}_{\nu} )  \right].
\end{eqnarray}

    The $a_1(k) \to \pi(p) \rho(q)$ decay width can be obtained as
\begin{equation}
\Gamma(a_1\rho\pi)=\dpf{|\vec{p}|}{12\pi m_{a_1}^2}\left[ 
      2f_{a_{1}\rho \pi}^2+\left(\dpf{E_{\rho}}{m_{\rho}}
      f_{a_{1}\rho\pi}+\dpf{m_{a_1}}{m_{\rho}}g_{a_{1}\rho\pi}
      |\vec{p}|^2\right)^2 \right] \, ,
\end{equation}  
where $\vec{p}$ is the momentum of the $\pi$ meson evaluated in the rest frame of
the $a_1$, and
\begin{eqnarray}
f_{a_{1}\rho\pi} & = & {g^{2} f_{\pi} \over Z_{\pi}}~\left[
2 c + {q^2 \over m_{a_1}^2}\right] + {\kappa_{6} g^{2} f_{\pi}
\over
m_{\rho}^2}~p \cdot q, \nonumber
\\
g_{a_{1} \rho \pi} & = & - {\kappa_{6} g^{2} f_{\pi} \over
m_{\rho}^2}\ .
\end{eqnarray}

The ratio of $D$-wave to $S$-wave amplitudes for this final state
is  
\begin{equation}
\dpf{D}{S}=-\dpf{\sqrt{2}\left[f_{a_1\rho\pi} (E_{\rho}-m_{\rho})
	+g_{a_1\rho\pi} m_{a_1}|\vec{q}|^2 \right]}{f_{a_{1}\rho\pi}
(E_{\rho}+2m_{\rho})+g_{a_1\rho\pi} m_{a_1}|\vec{q}|^2} \ .
\end{equation}

This chiral formulation admits a $a_1 \pi \gamma$ direct contact term and
as such, deviates from the ``traditional'' Vector Meson Dominance approach. 
Including this, the radiative decay width one obtains for $a_1 \to \pi
\gamma$ is

\begin{equation}
\Gamma(a_1 \pi \gamma)=\dpf{1}{24}\dpf{\alpha \kappa_6^2 g^2 f_{\pi}^2}
      {m_{\rho}^4}\dpf{\left(m_{a_1}^2-m_{\pi}^2 \right)^3}{m_{a_1}^3} \, .
\end{equation}
Here is an opportune place for a short digression on the Vector Meson 
Dominance model (VMD). A recent discussion of its different representations 
can be found in Ref.~\cite{vmddiss}. We refer the reader to this reference, and those therein, for details. In short, the vector meson-photon vertex can 
either be a constant or the electromagnetic field strength tensor can couple 
to the rho field strength tensor to yield a $q^2$-dependent vertex, where 
$q$ is here the photon invariant mass \cite{klz}. In the latter 
case, an additional 
direct photon contact term is then necessary as the vector meson-photon mixing 
vanishes for real photons. The approaches described in this work invoke both 
versions of VMD. We have implemented the appropriate one for 
each case and also 
verified that gauge invariance was verified in the electromagnetic sector. 

Returning to the Lagrangian under scrutiny, 
following Refs.~\cite{ko,carter} we choose the parameters 
$g, c, \kappa_{6}$, and $Z_{\pi}$ guided by phenomenological 
considerations.
Two choices of those numbers are displayed in Table~\ref{table1}. Parameter
set I was used previously to address dilepton production in the reaction we are
concerned with here \cite{kimko}. However, this choice produces a 
wrong sign for the
$D/S$ ratio and its  $\chi^2$ can be somewhat improved. This  
is achieved by parameter set II, see Table~\ref{table1}.
\begin{table}
\begin{center}
\begin{tabular}{|c|cc|c|}
\hline
 & I& II &  DATA   \\
$\ \ \ g$  & 5.04 & 4.95 &  \\
$\ \ \ c$ & $-0.12$ & 1.29 &  \\
$\ \ \ \kappa_6$ & 1.25 & $-1.9014$ & \\
$\ \ \ Z_{\pi}$ & 0.17 & 0.83 & \\
\hline
$\Gamma{(a_1\pi \gamma) }$& 0.572 & 1.171 & $0.640 \pm 0.246$ MeV \\
$\Gamma{(a_1\rho \pi)}$ &313.4 & 579.1 & $\sim 400 $ MeV \\
$D/S$ & 0.078 & $-0.168$ & $-0.09 \pm 0.03$ \\
$\Gamma{(\rho e^{+} e^{-})}$& fit & 7.01 & $6.77 \pm 0.32$ KeV  \\
$\ \ \ \chi^2$& 8.9 & 7.5 & \\
\hline
\end{tabular}
\caption{\small Two parameter sets for the model of Ref.~\protect\cite{ko}, and the
associated phenomenology.}\label{table1}
\end{center}
\end{table}

\subsection{ A $U(2)_L \times U(2)_R $ effective chiral Lagrangian }

	A  $U(2)_L \times U(2)_R $ chiral model for pseudoscalar,
vector, and axial-vector mesons has been proposed by B. A. Li
\cite{li} which produces a successful description of meson phenomenology. 
 In this  effective chiral theory, the $a_1 (k) \pi (p) \rho (q)$
coupling is described by the following interaction Lagrangian:
\begin{equation}
{\cal L}_{a_{1}\rho\pi}=
A\vec{a}^{\mu}\cdot(\vec{\rho}_{\mu}\times\vec{\pi})
+B\vec{a}^{\mu}\cdot(\vec{\rho}^{\, \nu}\times\partial_{\mu\nu}
\vec{\pi}) ,
\label{eq:la} 
\end{equation}
where
\begin{eqnarray}
&A&=\dpf{2}{f_{\pi}}\left(1-{1\over 2\pi^{2}g^{2}}\right)^{-{1\over
2}}
\left\{{F^{2}\over g^{2}}+\frac{m^{2}_{a1}}{2\pi^{2}g^{2}}
-{2c\over g}(p\cdot q + p\cdot k)-\frac{3}
{2\pi^{2}g^{2}}\left(1-{2c\over g}\right)p\cdot q\right\}
\nonumber \\  
&B&=-{2\over f_{\pi}}\left(1-{1\over 2\pi^{2}g^{2}}\right)^{-{1\over
2}}
{1\over 2\pi^{2}g^{2}}\left(1-{2c\over g}\right)\ ,
\label{eq:ab} 
\end{eqnarray}
and,
\begin{eqnarray}
&&c=\frac{f^{2}_{\pi}}{2gm^{2}_{\rho}},\\
&&F^{2}=\dpf{f_{\pi}^2}{1-\dpf{2c}{g}} .  
\end{eqnarray}

        In this model, $f_\pi = 0.186$ GeV and the particle masses are
taken as input. Fitting $g=0.35$ yields a good list of
light-meson empirical properties \cite{li}.\footnote{In Eq.(62) of
ref. \cite{li}, we argue that a factor of $\sqrt{2}$ is missing. For
$m_{a_1}=1230$ MeV, we then get $D/S=-0.0895$. The experimental value is
$-0.09 \pm 0.03 \pm 0.01$. } 

\subsection{Another effective chiral Lagrangian } 

	Yet another effective chiral Lagrangian for
pseudoscalar, vector, and axial-vector mesons can be derived  
\cite{gom}. In
this model, the $\pi$ meson is described through the nonlinear $\sigma$
model, and the $\rho$ and $a_1$ mesons are included 
as massive Yang-Mills fields of the chiral
symmetry. 
This scheme has been used to discuss photon
emission\cite{song} and dilepton production\cite{skg} from hot
hadronic matter. 

	The lowest order interaction term for $a_1 \rho \pi$ is given by a 
lengthy expression given in terms of the meson matrices \cite{gom}.
	For the physical $a_1 \to \rho \pi$ decay, the vertex function
leading to the decay width is
\begin{equation}
\Gamma_{\mu \nu}(a_1 \to \rho\pi) 
                =i(g_{a_1 \rho\pi}g_{\mu\nu}-h_{a_1 \rho\pi}q_{\mu}k_{\nu}) \, ,
\end{equation}
where, 
\begin{eqnarray}
g_{a_1 \rho\pi}&=&\dpf{g}{\sqrt{2}}\left[-\eta_1 q^2 
               +(\eta_1-\eta_2)k\cdot q \right] \, , \cr 
h_{a_1 \rho\pi}&=&\dpf{g}{\sqrt{2}}(\eta_1-\eta_2) \, ,
\end{eqnarray}
with
\begin{eqnarray}
\eta_1&=&\left({1-\sigma\over 1+\sigma}\right)^{1/2}\left(gF_\pi\over
2m_\rho^2
\right)+{4\xi Z^2\over F_\pi\sqrt{1+\sigma}},\cr
\eta_2&=&\left({1+\sigma\over 1-\sigma}\right)^{1/2}\left(gF_\pi\over
2m_\rho^2
\right)-{4\sigma\over gF_\pi\sqrt{1-\sigma^2}},\cr
{\rm and}\cr
Z^2&=& 1 - {{g^2 F_\pi^2}\over{4 m_\rho^2}}\ .
\end{eqnarray}

Here, $F_\pi\approx 135$ MeV. 
     With the two sets of parameters which give
the $a_1 \to \rho \pi$ decay width $\Gamma(a_1 \to \rho \pi) \approx
400$ MeV \cite{song}, we further obtained the $D/S$ ratio for 
this effective Lagrangian: 
\begin{eqnarray}
{\rm set\ \ I} & :& g=10.3063,\ \  \sigma=0.3405, \ \ \xi=0.4473, \ \ 
                    D/S=0.357  \cr 
{\rm set\ \ II} & :& g=6.4483, \ \  \sigma=-0.2913, \ \  \xi=0.0585, \ \
                     D/S=-0.099 \ .
\label{eq:par}  
\end{eqnarray}

In this work, we shall use parameter set II. 
	The electromagnetic interaction is introduced by
imposing the $U(1)_{em}$ gauge symmetry on the effective chiral
Lagrangian. 

\subsection{Two more simple effective Lagrangians for $a_1 \rho
\pi$ interactions }

        Xiong, Shuryak and Brown \cite{xiong} have defined an effective
Lagrangian for $a_1 \rho \pi$ interactions, in order to calculate photon
production from hot hadronic matter.
This Lagrangian is
\begin{equation}
{\cal L}_{a_1 \rho \pi}=G_{\rho}a^{\mu}(p\cdot q g_{\mu\nu}
         -q_\mu p_\nu )\rho^{\nu}\pi .
\end{equation}
        The coupling constant $G_{\rho}$ is determined by fitting the
$a_1 \to \rho \pi$ decay width. Unfortunately, we can not reproduce the
numerical results of Ref.~\cite{xiong}. Using $m_{a_1}=1230$ MeV and
fitting the total decay width, we get
$G_{\rho}=11.425$ GeV$^{-1}$. Using the Vector Dominance Model
we obtain the coupling constant of the $a_1 \pi \gamma$ interaction as
$G_{\gamma}=0.573$ GeV$^{-1}$. This then implies a value of 
$\Gamma_{a_1 \to \pi \gamma}=1.94$ MeV, which is somewhat larger than
the experimental measurement of $0.640 \pm 0.246$ MeV. This effective Lagrangian
also predicts 
$D/S = 0.185$.   
\begin{table}[!h]
\begin{center}
\begin{tabular}{|c|cc|c|c|c|c|c|}
\hline
Source: & \protect\cite{ko} &   & \protect\cite{gom}   & \protect\cite{li} 
& \protect\cite{xiong} & \protect\cite{jhs} &  DATA   \\  
   &  I & II &  &  & & & \\ 
\hline     
$\ \ \ \Gamma{(a_1\rho \pi)}$ &313.4 & 579.1 & fit & 331.7 & fit & fit &
$\sim 400 $ MeV \\
$\ \ \ \Gamma{(a_1\pi \gamma) }$& 0.572 & 1.171 & 0.067 & 0.331 & 1.940 &
0.312 & $0.640 \pm 0.246$ MeV \\
$\ \ \ D/S$ & 0.078 & $-0.168$ & $-0.099$ & $-0.161 $ & 0.185 &  
0.045 &  $-0.09 \pm 0.03$ \\
$\chi^2$ & 11.8 & 9.6 & 1.8 & 3.1 & 37.3 & 7.3
  &\\
\hline
\end{tabular} 
\caption{\small A comparison of hadronic properties for  the 
interactions discussed in this
work. Note that  the $\chi^2$'s appearing in this Table were calculated using 
the experimental measurements that appear on it only. When different
parameter sets are involved for a given interaction, the distinction is
explained in the text. }\label{table2}
\end{center}
\end{table}

	The other simple effective Lagrangian which considers the $a_1 \rho \pi$
interaction was used to formulate a meson-exchange model for 
$\pi \rho$ scattering by Janssen,
Holinde and Speth\cite{jhs}. It is 
\begin{equation}
{\cal L}_{a_1 \rho \pi}=g_{a_1}(\partial_\mu \vec{\rho}_\nu 
          -\partial_\nu \vec{\rho}_\mu) \cdot [\vec{\pi} \times 
          (\partial^\mu \vec{a}^\nu-\partial^\nu \vec{a}^\mu)] .  
\end{equation}
	For $m_{a_1}=1230$ MeV, the coupling constant $g_{a_1}=2.285$
GeV$^{-1}$ follows from the usual fit of the $a_1$ decay width.
The $D/S$ ratio here is $D/S = 0.045$. 

A comparative summary of the different on-shell properties and predictions
is shown in Table \ref{table2}.

\section{Results for dilepton production from hot  hadronic matter }

\subsection{Effective Lagrangian approach}
In this section we shall compute rates of lepton pair emission with the
effective interactions introduced earlier. In this study we rely on
relativistic kinetic theory  to provide an idealized  dynamical framework. 
Because we are mainly interested in comparisons between different models, 
this line of reasoning is
entirely adequate. We also set any chemical potential to zero, for
simplicity. The temperature chosen for our baseline study is  T~=~150 MeV. 

	In general, the dilepton production rate from the annihilation of
two particles $a (p_1) $ and $b (p_2) $ can be written as\cite{gale} 
\begin{equation}
\dpf{dN}{d^4x dM^2}={\cal N}\int{d s} \int{d^3p_1\over(2\pi)^3}
                     {d^3p_2\over(2\pi)^3}f_a(p_1)f_b(p_2)  
                     \dpf{d \sigma(ab \to l^+l^-)}{d M^2} 
                    v_{\rm rel}\delta\left(s-(p_1+p_2)^2\right) ,
\end{equation}
where ${\cal N}$ is an overall degeneracy factor, $f(p)$ is the
distribution
function of the incoming particles at temperature $T$, $v_{\rm rel}$ is
the
relative velocity of the two particles, 
\begin{equation}
v_{\rm rel}=\dpf{\sqrt{(p_1\cdot p_2)^2-m_a^2 m_b^2}}{E_aE_b} ,
\end{equation}
 and $\sigma$ is the
dilepton production cross section for
the reaction $a+b \to l^++l^-$. Using Boltzmann distributions, one performs
the phase space integrations and obtains 
\begin{equation}
\dpf{dN}{d^4x dM^2}={\cal N}\dpf{T}{32\pi^4 M} K_1(M/T) 
                     \lambda(s, m_{a_1}^2, m_{\pi}^2)\sigma(M) , 
\end{equation}
where $K_1$ is a modified Bessel function, and $\lambda $ is the
kinematic triangle function
\begin{equation}
\lambda(x,y, z)=x^2-2x(y+z)+(y-z)^2 .
\end{equation}
 
The cross section $\sigma(M)$ depends only on the square of the invariant mass
of the dilepton $s=(p_1+p_2)^2=M^2$ and is easily related to the square of  
a spin-averaged scattering amplitude. This in turn can be written as
\begin{equation}
\left| \overline{\cal M} \right|^2=4\left(\dpf{4\pi \alpha}{s}\right)^2
L_{\mu \nu}H^{\mu \nu},  
\end{equation}
 	Where, $\alpha$ is the fine structure constant, and $L^{\mu \nu}$ 
is a leptonic tensor given by
\begin{equation}
L^{\mu \nu}=q_1^{\mu} q_2^{\nu} +q_2^{\mu}q_1^{\nu}-g^{\mu\nu}q_1\cdot q_2, 
\end{equation} 
and $H^{\mu \nu}$ is a hadronic tensor for the process. Note that $ (q_1 +
q_2 )^2 = s = M^2$. The hadronic tensor is calculated from the meson vertex
function of the appropriate Feynman diagram with the relevant Lagrangian. 

\begin{figure}
\begin{center}
\includegraphics[angle=90, width=10cm]{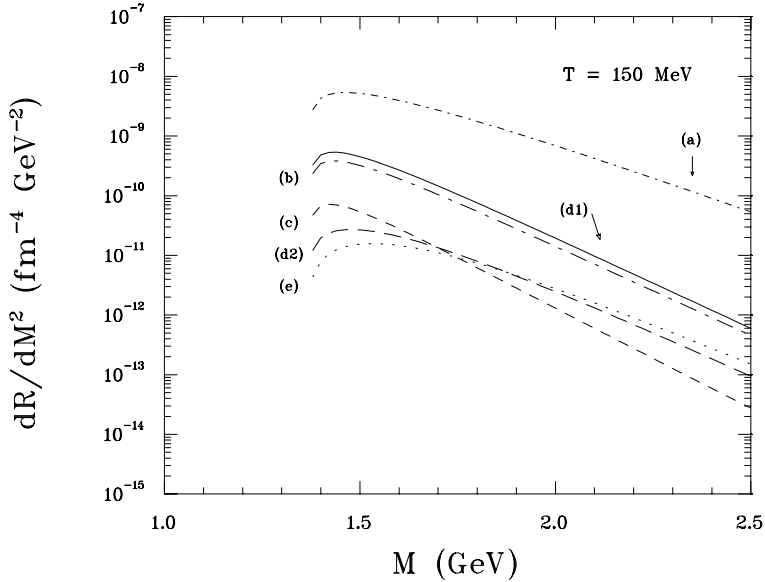}
\end{center}
\caption{\small  
Dilepton production rates from the $\pi + a_1 \to e^+ e^-$
reaction, at $T$=150 MeV.  The curves labels are as follows: 
 (a) Ref.~\protect\cite{li}, (b) Ref.~\protect\cite{gom}, (c) 
Ref.~\protect\cite{jhs}, (d1) and
(d2) Ref.~\protect\cite{ko} parameter sets I and II (see Table \ref{table1}), (e)
Ref.~\protect\cite{xiong}.}
\label{result150}
\end{figure}

We plot in Fig.~\ref{result150} the rates for dielectron production from a gas
of hot mesons, at a temperature of T = 150 MeV. We concentrate 
on $\pi a_1$ reactions. All of the different interactions enumerated
previously have been considered and their contribution appears here. 
A striking feature 
is that the rates calculated with those Lagrangians span
two-and-a-half orders of magnitude. Recall that all of them (except one) 
have been used in
the literature to perform dilepton calculations very similar to the one
done here. The features illustrated in
Fig.~\ref{result150} are essentially temperature-independent. We have verified
this by also performing calculations at a temperature of T = 100 MeV. 
We conclude from this that 
off-shell effects are indeed quite
important. Considering
the curves labeled (d1) and (d2), one can verify by looking again at Table
\ref{table2} that a modest change in the $\chi^2$ calculated on shell can
result in an important variation in off-shell behaviour. This tells us that
although the on-shell $\chi^2$ can be used as a goodness-of-fit parameter,
it is far from being enough to specify unambiguously which interaction to
use in situations like the one at hand. One has to proceed with caution.

We now turn to an approach which should have the off-shell effects buit-in,
in order to help us deciding upon an interaction to use. 
 
\subsection{Using experimentally-constrained spectral functions}

Above, we have found large differences in lepton producing rates using different
hadronic Lagrangians to model the hard vertices. This result is worrisome,
as many of those interactions have been used in the past for dilepton
calculations, and there is no doubt that they will be used again in the
future.
We then turn to an alternative approach, with the hope
that this will assist us in the selection of an appropriate theory. In the
context of heavy ion reactions, it has been argued recently that lepton
pair production cross sections could in fact be determined from the
spectral functions extracted from the inverse process: $e^+ + e^- \to {\rm
hadrons}$ \cite{huang95}. We refer the reader to the appropriate 
reference for the complete details. It will suffice here to state that the
rate for the emission of dilepton pairs of invariant mass $M$, at
temperature $T$ in a given reaction is given by
\begin{eqnarray}
\frac{d R}{d M^2}\ = \ \frac{4 \alpha^2}{2 \pi} M T K_1 ( M / T )\
\rho^{\rm em} ( M )\ ,
\end{eqnarray} 
where $\alpha$ is the usual fine structure constant, $K_1 ( x )$ is a 
modified Bessel function, and  $\rho^{\rm em} ( M )$
is a zero-temperature spectral function. This quantity is extracted from
$e^+ e^-$ annihilation data through 
\begin{eqnarray}
\rho^{\rm em} ( s )\ = \ \frac{s\, \sigma (e^+ e^- \to {\rm hadrons} )}{16
\pi^3 \alpha^2}\ . 
\label{rho}
\end{eqnarray}
We then follow the same approach as that of a current algebra calculation
\cite{pen80} to extract the $\pi a_1$ contribution from $e^+ e^- \to 4
\pi^\pm$ data. One can then obtain $\sigma (e^+ e^- \to \pi a_1 )$,
which can then be used in Eq. (\ref{rho}) to extract the spectral function. 
\begin{figure}[!h]
\begin{center}
\includegraphics[angle=90, width=10cm]{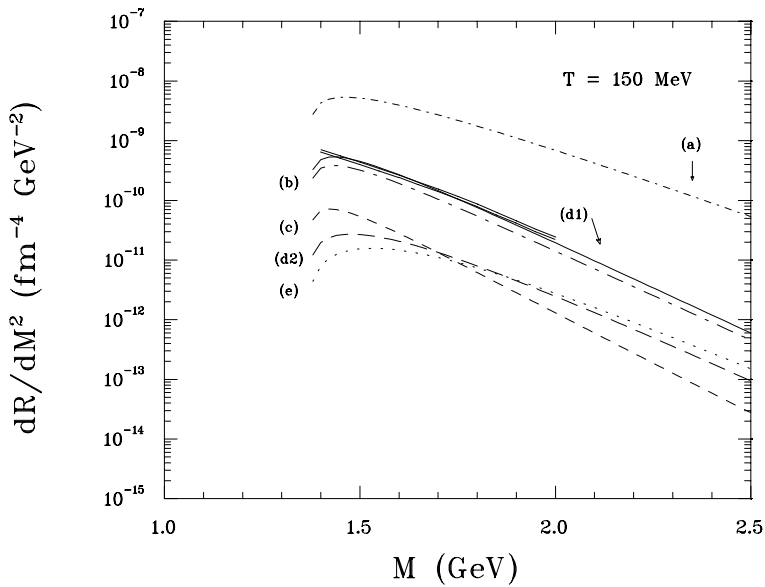}
\end{center}
\caption{\small  
The curve labels are the same as for the previous two figures. The
double line represent the rates derived from cross sections calculated with
the methods of Refs.~\protect\cite{huang95,pen80}.}
\label{totcomp150}
\end{figure}
Note that this procedure is straightforward and relatively free from
ambiguity, numerical or otherwise. It is tantamount to a direct evaluation
of the dilepton cross sections by the same theoretical methods.

We plot on Fig.~\ref{totcomp150} the rates previously shown on
Fig.~\ref{result150}, accompanied by the rate evaluated with the method
outlined above. Note that this latter calculation does not run over the entire
invariant mass range covered in the plot, as we have chosen not to stretch
the validity of the theoretical approach which was deemed optimal in the
invariant mass range covered by the double line \cite{pen80}. This coverage
is sufficient for us to make our point. Considering that the spectral
function is constrained by data, we use this analysis as an extra tool to
discriminate between the various interactions considered in this work. We
observe that the spectral function results are overshot by the interaction
from Ref.~\cite{li}.  They are underpredicted by the rates related 
to the Lagrangians in Refs. \cite{xiong} and \cite{jhs}, 
and by rates obtained  with parameter set II of the
interaction from Ref.~\cite{ko}. The interactions from  
Refs.~\cite{gom} and \cite{ko} (parameter set I) yield results in 
good agreement with the spectral function determination. Even though both 
Lagrangians seem to exhibit a satisfactory off-shell behaviour, 
going back
to Table~\ref{table2} one observes that the one from 
Ref.~\cite{gom} also produces an excellent $\chi^2$ for
on-shell hadronic properties. It appears that, with the parameters
used in this study and considering both on-shell and off-shell behaviours, 
this interaction achieves the compromise we were seeking. It is also 
a satisfying one from a theoretical point of view \cite{ecker}.  
Note in closing that 
the rates above were determined in the narrow-width approximation for the 
$a_1$. While at first sight this limit seems unreasonable, it has been shown
to affect little the magnitude of the thermal rates \cite{songko}. 
Its main effect is to 
soften the threshold of this specific reaction. 

\section{Conclusion}
 
	We have calculated dilepton emission from $\pi a_1$ reactions at
finite temperature using several different Lagrangian found in the
literature. We have found widely different results. One might argue that
some of those interactions were derived for entirely different purposes,
therefore comparing them on the basis of dilepton emission seems vaguely
unappropriate. One should however remember that all of the Lagrangians
(except that of Ref.~\cite{li}) {\em have} been used previously in such
exercises. Up to now, a critical comparison of their results was lacking. 
With the help of an experimentally-constrained spectral function, combined
with  a quantitative analysis of on-shell properties, we were able to
select an adequate interaction. 
It is clear that a companion study to this one will consider the rates for
photon emission. This work is in progress and will be reported on later.
There are however indications that for photons, the differences arising
from the use of different Lagrangians will be
less severe \cite{kimko}. This probably owes to the fact that the mass
shell condition for real photons has a restraining effect.

It should be clear that our goal was not an exhaustive survey of the
parameter space relevant to each of the models we have discussed. We 
viewed these interactions as representative of what is currently on the market. 
One could in principle devise a new completely phenomenological interaction.
Our study highlights the important issue of what 
needs to be done in order to give credibility to
results obtained in an environment as potentially complex as that of 
high energy heavy ion collisions.

\section{Acknowledgements}
Our research is supported in part by the Natural Sciences and
Engineering Research Council of Canada, in part by the Fonds FCAR of
the Qu\'ebec Government, and in part by the US Department of Energy under grant number DOE/DE-FG02-88ER-40388. One of us (C. G.) is happy to acknowledge 
discussions with, and the hospitality of, the Nuclear Theory Group of the
State University of New York at Stony Brook, where this work was completed.
We are also happy to acknowledge useful discussions with S. Rudaz.

\newpage

\end{document}